\documentclass[12pt,preprint]{aastex}
\slugcomment{}

\begin{document}
\title{THE ORIGIN OF THE PLATEAU AND LATE REBRIGHTENING IN THE AFTERGLOW OF GRB 120326A}
\author{S. J. Hou\altaffilmark{1,2}, J. J. Geng\altaffilmark{3}, K. Wang\altaffilmark{3}, X. F. Wu\altaffilmark{2}, Y. F. Huang\altaffilmark{3},\\ Z. G. Dai\altaffilmark{3}, and J. F. Lu\altaffilmark{1}}
\altaffiltext{1}{Department of Astronomy and Institute of Theoretical Physics and Astrophysics, Xiamen University, Xiamen, Fujian 361005}
\altaffiltext{2}{Purple Mountain Observatory, Chinese Academy of Sciences, Nanjing 210008, China; xfwu@pmo.ac.cn}
\altaffiltext{3}{School of Astronomy and Space Science, Nanjing University, Nanjing 210093, China}

\begin{abstract}
GRB 120326A is an unusual gamma-ray burst (GRB) which has a quite long plateau and a very late rebrightening both in X-ray and optical bands. The similar behavior of the optical and X-ray light curves suggests that they maybe have a common origin. The long plateau starts from several hundred seconds and ends at tens of thousands seconds. The peak time of the late rebrightening is about 30000 s. We analyze the energy injection model by means of numerical and analytical solutions, considering both the wind environment and ISM environment for GRB afterglows. We especially study the influence of the injection starting time, ending time, stellar wind density (or density of the circumburst environment), and injection luminosity on the shape of the afterglow light curves, respectively. We find that the light curve is largely affected by the parameters in the wind model. There is a ``bump'' at the late time only in the wind model too. In the wind case, it is interesting that the longer the energy injected, the more obvious the rebrightening will be. We also find the peak time of bump is determined by the stellar wind density. We use the late continuous injection model to interpret the unusual afterglow of GRB 120326A. The model can well fit the observational data, however, we find that the time scale of the injection must be larger than ten thousands seconds. This implies that the time scale of the central engine activity must be more than ten thousands seconds. This can give useful constraints on the central engine of GRBs. We consider a new born millisecond pulsar with strong magnetic field as the central engine. On the other hand, our results suggest that the circumburst environment of GRB 120326A is very likely a stellar wind.
\end{abstract}


\section{INTRODUCTION\label{sec:intro}}
Gamma-ray bursts (GRBs) are violent phenomena in the Universe, which radiate tremendous energy about $10^{51} - 10^{54}$ ergs from fractions of a second to tens of seconds. The widely accepted model is the fireball shock model (Goodman 1986; Rees \& M\'{e}sz\'{a}ros 1992, 1994;  M\'{e}sz\'{a}ros \& Rees 1992; Piran 1999). In this model, it is believed that the electrons are accelerated by internal shocks or external shocks. Internal shocks are likely developed near the site of optically thin fireball (Rees \& M\'{e}sz\'{a}ros 1994), which give birth to the prompt emission due to collisions of shells with each other. Soon after the burst, the relativistic ejecta continue to spread out to form an external shock. The external shock sweeps up interstellar medium (Blandford \& McKee 1976; Piran et al. 1993; Sari et al. 1998), which is believed to produce X-ray, optical/IR and radio emission, i.e. the afterglow.

The {\it Swift} satellite (Gehrels et al. 2004), which was launched by NASA on November 20, 2004, opened a new era for GRBs researches. The X-ray Telescope (XRT) (Burrows et al. 2005b) onboard the {\it Swift} has detected several hundreds of X-ray afterglows and the results are fruitful (M\'{e}z\'{a}ros 2006; Zhang 2007). In the summarization of the X-ray afterglow data, a canonical X-ray afterglow light curve includes five components (Zhang et al. 2006; Nousek et al. 2006). These five components include the steep decay phase (Zhang et al. 2007; Zhang et al. 2009), the shallow decay phase (Liang et al. 2007), the normal decay phase (Willingale et al. 2007), the post-jet break phase (Liang et al. 2008) and X-ray flares (Burrows et al. 2005a; Dai et al. 2006), respectively. For many bursts, we do not observe all components simply because of inadequate observations. Lots of samples only consist of single-power component (Liang et al. 2009). Recently, Zhang et al. (2013) studied the long-term central engine activities in the X-ray afterglow. Hou et al. (2014) studied a special sample, GRB 130925A, which appeared to a series of flares in the X-ray afterglow. Different samples show different temporal structures, increasing our perplexity. Meanwhile, Li et al. (2012) had systematically decomposed the optical afterglow light curves, indicating that the structures and composition of the optical afterglows were more complex than the X-ray afterglows.

For X-ray afterglows, the shallow decay phase is still a puzzle (Zhang 2007). The broadband afterglows, which usually decay as a power-law function of time with an index of $\alpha \sim 1.2$ (normal decay phase), are believed to be associated with the external shock. If the external shocks are refreshed by continuous energy injection into the blast wave, a shallow decay phase prior to the normal decay phase could be observed. The shallow decay may be due to the following mechanisms: (i) Energy injection invoking a long-term central engine (Dai \& Lu 1998; Zhang \& M\'{e}sz\'{a}ros 2001; Zhang et al. 2006; Nousek et al. 2006), (ii) Late internal shock model ( Zhang \& M\'{e}sz\'{a}ros 2002; Zou et al. 2013), (iii) Off-axis jet model (Eichler \& Granot 2006; Toma et al. 2006), (iv) Central engine model (Kumar et al. 2008; Geng et al. 2013 ). To determine which one of these models is correct, we need more observational data.

A quite long plateau followed by a very late rebrightening is observed in the afterglow of GRB 120326A in X-ray and optical bands, which is observed for the first time. The feature of rebrightening behavior around 30000 s is not easy to understand. The phenomenon is so special that a lot of telescopes observe this burst later. We collect a lot of observational data, including early and late afterglow in optical band. It can potentially helps us to understand the underlying mechanism of this afterglow. Unlike the common X-ray flares or the prompt pulses, which are usually characterized by a rapid rise and an exponent decay (Kocevski et al. 2007; Norris et al. 2005; Chincarini et al.2007, 2010; Margutti et al. 2011; Li et al. 2012), GRB 120326A shows a slight bump overlapping the afterglow light curve.

In this work, we mainly use the energy injection model in the wind case to explain the particular phenomenon of GRB 120326A. The plan of the paper is the following: in Section 2, we briefly describe the observation data. We review the energy injection model and fit the X-ray data in Section 3. In Section 4, we present our conclusions and discussion. The cosmological parameters $H_{0}$ = 71 km $\rm s^{-1}$ $\rm Mpc^{-1}$, $\Omega_{M}$ =0.3, and $\Omega_{\Lambda}$ = 0.7 have been adopted throughout our study.

\section{DATA ANALYSIS\label{sec:data}}

\subsection{Prompt Emission}
GRB 120326A was first detected at $T_0=$01:20:29 UT on 2012 March 26 by the Burst Alert Telescope (BAT) onboard the {\it Swift} satellite and was located at a position of $\alpha$ = $18^{h} 15^{m} 42^{s}$, $\delta$ = +69$^{\circ}$15$'$37$''$.0 (J2000), with a 90$\% $ containment radius of 4$'$.1 (Siegel et al. 2012). It was also triggered and located by the Fermi Gamma-ray Burst Monitor (GBM) (Collazzi 2012). The light curve of prompt emission was a single pulse with a duration $(T_{\rm 90})$ of about 12 s (50-300 keV)(Barthelmy et al. 2012), showed in left panel of Figure 1. The redshift of GRB 120326A was 1.798 (Kruehler et al. 2012; Tello et al. 2012).

We process the Fermi/GBM data using RMFIT. The time-averaged spectrum from $T_{0}$ -3.58 s to $T_{0}$ +13.82 s, as shown in the right panel of Figure 1, is fitted well by the Band function, yielding a relatively low peak energy $E_{p}= 64.42 \pm 7.54$ keV, a typical low energy photon index $\alpha = -1.18 \pm 0.15$ and a quite soft high energy photon index $\beta= -3.04 \pm 1.06$. The high energy photon index is not confined very well. The reduced chi-squared of the fit is $\chi^2 = 239.8/211 = 1.14$. The total fluence of the prompt emission in the 10 - 1000 keV band is $3.54 \pm 0.17 \times 10^{-6}$ erg cm$^{-2}$, which corresponds to an isotropic energy release $E_{\rm iso} = 1.96 \pm0.17 \times10^{51}$ ergs. According to the empirical $\Gamma_{0}-E_{\rm iso}$ relationship (Liang et al. 2010), the initial Lorentz factor can be estimated to be $\Gamma_{0}\simeq120$.

\subsection{X-ray and Optical Afterglows}
The XRT started to observe GRB 120326A from 59.5 seconds after the BAT trigger (Kennea et al. 2012). The X-ray light curve of GRB120326A is showed in Figure 2, which is taken from the UK {\it Swift} Science Data Centre at the University of Leicester (Evans et al. 2007, 2009). The X-ray light curve has a steep decay, of which the decay index is about 3.4 and it lasts from 52 s to 268 s after the BAT trigger. Then there is a data gap until about $T_0+3700$ s due to the first earth occultation. A plateau emerges in the second orbit observation (or maybe before it) and ends at about 20 ks after the BAT trigger. After the plateau, the light curve shows a rebrightening which peaks at about 30 ks - 40 ks after the BAT trigger with the rising slope $\sim$ 2.22. It is not like the common X-ray flares or the prompt pulses. This phenomenon is very peculiar and it is the first time to be observed in the afterglow so obviously. We use the web-based (http://www.swift.ac.uk/) analysis system for the XRT data analysis (Evans et al. 2007, 2009). An average spectrum is obtained from 3700 s to 80 ks, during which the energy injection is thought to play the role. The spectrum can be fitted with an absorbed power-law with a photon index of $1.89 \pm 0.06$. The best-fit is achieved with the absorption column density $\rm 4.5 \pm 1.2 \times 10^{21} \ cm^{-2}$.

In spite of the very dim optical afterglow, it was observed by a lot of ground-based telescopes (e.g., Klotz et al. 2012a; Zhao et al. 2012). We collect the optical data from the Gamma-ray Coordination Network (GCN). Considering the different filters of these observations, we select the $R$ and $r$ bands data, which are showed in Figure 2 by black dots. By combining the data from the GCNs 13111, 13119 and 13192 which are unfiltered observations (Guidorzi 2012; Hentunen et al. 2012; Quadri et al. 2012), we can get a well-limited light curve of the optical emission, which are showed in Figure 2 by the black open circles. From Figure 2, we can see the optical afterglow light curve is composed of three parts: a decay from about 100 s to 1000s, a plateau from  about 1000 s to 10 ks (may be more longer), and a brightening from 10 ks to dozens of thousands seconds. The detailed optical data are listed in Table 1.

At early stage of afterglow, the X-ray shows a steep decay which is to be due to the curvature effect from the internal shock and the optical band shows a normal decay from the forward external shock. Optical and X-ray light curves in the late stage are very similar, though there may be a color evolution during the rebrightening (Kuin et al. 2012). Due to their similar shapes, they may be of the same origin.

\section{MODELING THE AFTERGLOW OF GRB 120326A}
The shallow decay phase is often observed in X-ray and optical afterglows (Liang et al. 2007). The temporal decay slope is about 0.5, which is flatter than the temporal slope of normal decay ($\thicksim$ 1.2). The slope of shallow decay cannot be explained by the standard afterglow model (M\'{e}sz\'{a}ros \& Rees 1997; Sari et al. 1998; Chevalier \& Li 2000; Sari \& Esin 2001). This phenomenon is difficult to understand. For some bursts, no spectral evolution is observed during the phase transition, which rules out the crossing of spectral break frequencies in the observing band (Zhang 2007). The energy injection model is still a preferred model to explain the shallow decay phase. In the framework of the fireball shock model, all the shells merge into a thick shell which continues to move forward and interact with the surrounding medium to form the external shock after the prompt emission. The external shock accelerates electrons to relativistic speed. So a fraction of shock energy will be transported to the swept-up medium as internal energy. The synchrotron radiation from the relativistic electrons contribute to the afterglow in X-ray, optical and radio bands. But after the prompt emission, a new born millisecond pulsar (or black hole) with strong magnetic field and rapid rotation can be born. It can produce a Poynting-flux-dominated wind (Dai \& Lu 1998). The strong Poynting flow can be injected directly into the external shock and its energy might be much larger than the initial energy of the external shock. This so-called energy injection process is used to interpret the plateau or ``bump'' features in X-ray and optical afterglow light curves.

\subsection{Shock Dynamics and Synchrotron Radiation}
A generic dynamical model of GRB outflows was proposed by Huang et al. (1999, 2000), and it had been widely used to calculate the afterglow light curves. Recently, the effects of some subtle factors such as the adiabatic pressure and radiative losses on the dynamics were further studied (van Eerten et al. 2010; Pe'er 2012; Nava et al. 2013). When Poynting-flux energy injection is taken into account, the basic equation for GRB outflow dynamics during the afterglow phase can be modified as (also see Kong \& Huang 2009; Liu et al. 2010)
\begin{equation}
\frac{d \gamma}{d m} = - \frac{(\gamma^{2} - 1)-\frac{1 - \beta}{\beta c^{3}} \Omega_{j} L(t - R/c)}{M_{ej} + 2 (1 - \varepsilon) \gamma m + \varepsilon m} \frac{d R}{d m},
\end{equation}
where $\beta$ and $\gamma = 1/\sqrt{1 - \beta^{2}}$ are the bulk velocity and Lorentz factor of the shocked medium, $m$ is the mass of the swept-up surrounding medium by the shock, $\Omega_{j} = (1 - \cos \theta_{j}) / 2$ is the beaming factor of the GRB outflows, $\theta_{j}$ is the half-opening angle of the jet, $M_{ej}$ is the initial mass of the jet, $\varepsilon$ is the radiative efficiency, $R$ is the radius, $c$ is the speed of light, and $L$ is the luminosity of the additional energy injection into the forward shock.

If the central engine is a magnetar, the Poynting flux power evolves with time as $L = L_{0} (1 + t / T)^{-2}$, where $L_{0}$ is the initial luminosity at $t = 0$, $T$ is the characteristic spin-down timescale. We assume the magnetar with initial spin period $P$, surface magnetic field strength $B$, moment of inertia $I$, radius $R_{M}$, and angle between the rotation axis and magnetic dipole moment $\theta$. Since the typical initial luminosity $L_{0}$ and timescale $T$ depend on the parameters of the magnetar (Dai 2004; Dai \& Liu 2012) as: $L_{0} = 4.0 \times 10^{47} B^{2}_{\bot,14} R^{6}_{M,6} P^{-4}_{-3}$ erg s$^{-1}$ and $T = 5.0 \times 10^4 B^{-2}_{\bot,14} R^{-6}_{M,6} P^{2}_{-3} I_{45}$ s, where $B_{\bot,14}$ = $B \sin \theta/(10^{14}$ G), $R_{M,6} = R_{M}/(10^{6}$ cm), $P_{-3} = P/(10^{-3}$ s) and $I_{45} = I/(10^{45}$ g $\rm cm^{2}$).

The afterglow photons mainly come from synchrotron radiation of electrons accelerated by the external shock (Sari et al. 1998; Sari \& Piran 1999a, 1999b; Gao et al. 2013). The electron distribution is assumed as: $n_{\gamma}=n_{0} \gamma^{-p} (\gamma_m \leq \gamma \leq \gamma_{max})$ after shock acceleration, where $p$ is the power law index of electron energy distribution, $\gamma_m=\epsilon_{e}\frac{p-2}{p-1}\frac{m_{p}}{m_{e}}(\gamma-1)$ is the minimum Lorentz factor of the electrons, $\gamma_{max}= (\frac{6\pi q_{e}}{\sigma_{T}B(1+Y)})^{\frac{1}{2}}$ is the maximum Lorentz factor of the electrons, $Y$ is the energy ratio between the inverse Compton component and the synchrotron component, $\epsilon_{e}$ is shock energy equipartition parameter for electrons, and $\sigma_{T}$ is the Thomson cross section. The cooling Lorentz factor of electrons is $\gamma_{c}=\frac{6\pi m_{e}c(1+z)}{\sigma_{T}\gamma B^{2}(1+Y)t}$.

In general, there are two types of medium surrounding the massive star: homogeneous interstellar medium (ISM) type and wind type. Liang et al. (2013) argued that the medium surrounding some GRBs evolved from the wind case to the ISM case at certain radius. However, Yi et al. (2013) suggested that the environment was neither a wind case nor a ISM case. For simplicity, we use the wind + energy injection model and the ISM + energy injection model to calculate the afterglow, respectively.

In the wind case, the typical synchrotron frequency, the cooling frequency and the maximum peak flux density are $\nu_{m} \approx 2 \times 10^{13} \epsilon_{B,-1}^{1/2} (\frac{1+z}{2})^{1/2} \bar{\epsilon}_{e,-1}^{2} E_{52}^{1/2} t_{\rm day}^{-3/2}$ Hz, $\nu_{c} \approx 8.9 \times 10^{11} (\frac{1+z}{2})^{- 3/2} \epsilon_{B,-1}^{- 3/2} E_{52}^{1/2} A_{\ast}^{-2} t_{\rm day}^{1/2}(1+Y)^{-2}$ Hz, and $F_{\nu, \rm max} = 2.3\times 10^{4}(\frac{1+z}{2})^{3/2} \epsilon_{B,-1}^{1/2} E_{52}^{1/2} A_{\ast} D_{L, 28}^{-2} t_{\rm day}^{-1/2}$ $\mu$Jy, where $\bar{\epsilon_{e}}=\epsilon_e (p-2)/(p-1)$, $\epsilon_{B}$ is shock energy equipartition parameter for magnetic fields, $A_{*}$ is the wind parameter (the number density of the wind is $n = 3 \times 10^{35} A_{*} r^{-2}$ cm$^{-3}$), $D_{L, 28}$ is the luminosity distance $D_{L}$ in units of $10^{28}$ cm, and $E_{52}$ is the initial isotropic kinetic energy $E_{\rm K, iso}$ in units of $\rm 10^{52}$ ergs. Taking the energy injection power as $L(t) \propto L_{0} t^{-q}$ (so $E\propto t^{1-q}$), then the typical synchrotron frequency scales as $\nu_{m} \propto t^{- (2 + q)/2}$, the synchrotron cooling frequency scales as $\nu_{c} \propto t^{(2 - q)/2} $, and the the peak flux density $F_{\nu,\rm max} \propto t^{-q/2}$ (Zhang et al. 2006). So, the synchrotron radiation flux density at the observing frequency $\nu$ (For simplicity, we only consider the optical and X-ray emission) is:
\begin{equation}
F_{\nu}=\cases{
(\nu/\nu_{c})^{-1/2}F_{\nu,\rm{max}}, &
$\nu_{c}<\nu<\nu_{m}$, \cr
(\nu/\nu_{m})^{-(p-1)/2}F_{\nu,\rm{max}}, &
$\nu_{m}<\nu<\nu_{c}$, \cr
\nu_m^{(p-1)/2}\nu_c^{1/2}\nu^{-p/2}F_{\nu,\rm{max}}, &
$\max\{\nu_{m},\nu_{c}\}<\nu<\nu_{\rm{max}}$.}
\end{equation}
As we can see, the evolution of $\nu_m$, $\nu_c$ and $F_{\nu,\rm{max}}$ actually determine the temporal evolution of the afterglow light curve.

In the ISM case,  the typical synchrotron frequency, the cooling frequency and the maximum peak flux density are $\nu_{m} \approx 1 \times 10^{13} \epsilon_{B,-1}^{1/2} (\frac{1+z}{2})^{1/2} \bar{\epsilon}_{e,-1}^{2} E_{52}^{1/2} t_{\rm day}^{-3/2}$ Hz, $\nu_{c} \approx 8.2 \times 10^{11} (\frac{1+z}{2})^{- 1/2} \epsilon_{B,-1}^{- 3/2} E_{52}^{- 1/2} n_{0}^{-1} t_{\rm day}^{- 1/2}(1+Y)^{-2}$ Hz, and $F_{\nu, \rm max} = 8.2\times 10^{4}(\frac{1+z}{2}) \epsilon_{B,-1}^{1/2} E_{52} n_{0}^{1/2} D_{L, 28}^{-2}$ $\mu$Jy, respectively, where $n_{0}$ is the density of the circumburst environment. Assuming the energy injection power as $L(t) \propto L_{0} t^{-q}$, then the typical synchrotron frequency scales as $\nu_{m} \propto t^{- (2 + q)/2}$, the synchrotron cooling frequency scales as $\nu_{c} \propto t^{(q - 2)/2} $, and the the peak flux density scales as $F_{\nu, \rm max} \propto t^{1 - q}$ (Zhang et al. 2006). Here, the synchrotron radiation flux density at the observing frequency $\nu$ can also be described by Equation (2).

\subsection{Parameter Effects of the Energy Injection Model}
The energy injection should be carried out within a period of time, starting from $T_{\rm start}$ and ending at $T_{\rm end}$. For simplicity, we only consider a constant injection luminosity, i.e.,
\begin{equation}
L(t) \sim L_{0},~~~~~~~T_{\rm start}\leq t \leq T_{\rm end}.
\end{equation}
Combining Eq.(4) and Eq.(1), we can calculate the evolution of the external shock subjecting to the energy injection from a strongly magnetized millisecond pulsar. Following the procedure described in Huang et al. (2000), we can calculate both X-ray and optical afterglows using any set of model parameter values. Besides, our numerical code has also included the effect of equal arrival time surface (EATS, see Huang et al. 2007) and the effect of synchrotron self-absorption by electrons which might be important for optical emission during the early phase (Wu et al. 2003).

Here, we respectively analyze the afterglow light curves in the wind and ISM model:

\emph{In the wind model:} We investigate the effects of the wind parameter $A_{*}$, the starting time $T_{\rm start}$ and ending time $T_{\rm end}$ of energy injection, and the injection luminosity $L_{0}$ on the light curves of the afterglow through numerical calculations. To explore the effect of one of the above model parameters, we fix the values of the other 3 parameters. In our calculations, the standard choice of the values of these parameters are $A_{*}=0.1$, $T_{\rm start}=100$ s, $T_{\rm end}=10$ ks (the only exception is when investigating the effect of $A_{*}$, $T_{\rm end}=30$ ks is adopted), and $L_0=1.87\times 10^{49}$ erg s$^{-1}$. The typical values adopted for the remaining model parameters of the afterglow are $\Gamma_{0}=300$, $E_{\rm K, iso}=2.0\times10^{51}$ ergs, $\theta_{j}=0.05$ rad, $p=2.3$, $\epsilon_{B}=0.01$, and $\epsilon_{e}=0.1$, where $\Gamma_{0}$ is the initial Lorentz factor of the jet. With these parameters, we calculate the afterglow light curves under different conditions, as shown in Figures 3 and 4.

We briefly describe the effects of the different parameters on the X-ray light curves. Figure 3a shows the light curves for different wind parameter $A_{*}$. We set the $T_{\rm end} =$ 30 ks. When $A_{*}$ is 0.05, the peak time of the ``bump'' is about 3 ks, while when $A_{*}$ is 0.2, the peak time of the ``bump'' is about 10 ks. For a smaller $A_{*}$, the peak time is earlier and the peak flux is larger. In addition, the peak time is always less than $T_{\rm end}$. When $A_{*}$ is small enough, there will be a plateau/shallow decay after the peak of the ``bump''(e.g., $A_{*}$ = 0.1). When $A_{*}$ is large enough, there will be a shallow decay prior to the peak (e.g., $A_{*}=0.4$). These phenomena reveal that the wind parameter $A_{*}$ determines the peak time of the ``bump''. It also proves that the medium density plays a very important role in shaping the afterglow light curves in the energy injection model. The calculated light curve showing early shallow decay and late narrow ``bump''/rebrightening is a novel prediction by the energy injection model with a wind-type environment. As we will show in the next subsection, such a specific model can interpret the peculiar X-ray and optical afterglows of GRB 120326A quite well.

$T_{\rm end}$ affects the duration of the plateau/shallow decay as well as the late-time X-ray flux. When $T_{\rm end}$ is larger than the peak time which is determined by $A_{*}$, the light curve will show a ``shallow decay'' or ``plateau'' from the peak time to $T_{\rm end}$, as shown in Figure 3b. This is understandable because a larger $T_{\rm end}$ corresponds to more energy injected into the external shock. Figure 3c shows to the effect of the energy injection starting time $T_{\rm start}$. We find that the later the energy injection begins, the more obvious the rebrightening will be. This is in fact the zero-time effect. We also note that the $T_{\rm start}$ affects the peak time and the peak flux of the ``bump''. The later $T_{\rm start}$ is (the less the energy is injected into the external shock), the later the peak time will be, and the smaller the peak flux will be. Figure 3d shows the effect of the energy injection luminosity $L_{0}$. From the figure 3d, we can see that an obvious rebrightening is positively correlated with the $L_{0}$. The larger the injection luminosity is, the more obvious the rebrightening can be. The parameter effects on the optical afterglow light curves (Fig. 4) are quite similar to those on the corresponding X-ray light curves.

\emph{In the ISM model: }  Similar to the wind model, we study the effects of the density of the circumburst environment $n_{0}$, the starting time $T_{\rm start}$ and ending time $T_{\rm end}$ of energy injection, and the injection luminosity $L_{0}$ on the light curves of the afterglow through numerical calculations too. When we study the effect of one of the above model parameters, we take the standard parameters as $n_{0}=1.0$ $\rm cm^{-3}$, $T_{\rm start}=100$ s, $T_{\rm end}=30$ ks, and $L_0=1.87\times 10^{49}$ erg s$^{-1}$, respectively. We find that although the light curve is also affected by these parameters, there is not a ``bump'' at the late time in the light curve which is like the shape of GRR 120326A (see Fig. 5 and 6, corresponding to X-ray and optical light curves, respectively). In same case, when the starting time $T_{\rm start}$ is relatively later and the injection luminosity $L_{0}$ is relatively larger, there is also a wide bump. But it is very similar to GRB 121027A (Wu et al. 2013) and is different from GRB 120326A (see Fig. 5c). More detailed analysis can also be found in Geng et al. (2013).

\subsection{Fitting to the Afterglow of GRB 120326A}
The long plateau and a very late rebrightening of GRB 120326A in X-ray and optical bands at the same time are almost an ideal template for testing the energy injection + wind model. This case may help us to reveal the details of the underlying energy injection and circum-burst environment. Note that we only consider the forward shock emission as in our modeling and assume the injection energy is purely in the form of Poynting flux.

 For the normal afterglows in the non-injection cases, the model parameters can be roughly estimated from the analytical results according to the multi-band observations (e.g., Liu et al. 2013). However, in GRB 120326A, it is unable to constrain all the parameters analytically from the light curves only in two bands. In fact, the energy injection process will even make the justification more difficult since the evolution of $\nu_{c}$ and $\nu_{m}$ also depends on the injection luminosity. In our model, there are ten parameters. We set some parameters as the typical values (Freedman et al. 2001; Wu et al. 2003): $p=2.2$, $\epsilon_{e}=0.1$, $\epsilon_{B}=0.01$. $p=2.2$ is consistent with the photon index of 1.89 of the spectrum from 3700 s to 80 ks if $\nu_{X} > \nu_{\rm opt} > \nu_{m} > \nu_{c}$. We set the half-opening angle a typical value too: $\theta_{j}$ = 0.05 rad (Frail et al. 2001; Lu et al. 2012). $E_{\rm K,iso}$ and $\Gamma_{0}$ are estimated from Section 2.1. The plateau of the optical light curve indicates that the injection starting time is around 600 s though the data seems scattering, thus we set $T_{\rm start} \simeq 600$ s. On the basis of the rebrightening of optical and X-ray light curve, we set $T_{\rm end}$ $\simeq$ 30 ks.  When $A_{*}$ is large enough, there will be a shallow decay prior to the peak. These phenomena reveal that the wind parameter $A_{*}$ determines the peak time of the ``bump'', which is discussed in Section 3.2. GRB 120326A is just an example of this case. Through the $T_{\rm end}$ $\simeq$ 30 ks, we can get the $A_{*}$ $\geq$ 0.4. Then there are only two parameter $L_{0}$ and $A_{*}$, which need to be determined by fitting the observations. There is a relatively large scattering in the plateau phase of the optical data, which makes it difficult for us to search the best parameters during the parameter space. Note that we want to address in this article is the origin of the plateau and rebrightening in GRB 120326A. After some trials, the rough fitting result is shown in Figure 2, in which $L_0=1.70\times10^{49}$ erg s$^{-1}$ and $A_{*}$ = 0.45 are adopted. Two points should be emphasized in the fitting process: First, we only consider the $R$ and $r$ bands optical data and the unfiltered band data are only used for reference. Second, we are mainly concerned with the plateau and rebrightening stage.

Figure 7 shows the evolution of $\nu_{c}$, $\nu_{m}$ and $\gamma$ with time. The left panel of Figure 7 shows that the evolution of these two frequencies differs slightly from the analytical ones since there exists a transition period from the non-injection case to the full injection case. This kind of evolution rests with the evolution of $\gamma$ given in the right panel of Figure 7. This difference finally results in the small bump during the energy injection. The same reason will also hold for the rising segment in the optical afterglow. Note that there is a steep rise for $\nu_{c}$ after $30000 s$, it is caused by synchrotron self-Compton (SSC) scattering between the photons and electrons and will have an obvious effect near $\nu_{c}$ = $\nu_{m}$ (Sari \& Esin 2001). However, this steep rise could not be responsible for the observed bump since it is after the peak time ($\sim 30000$ s).

\section{DISCUSSION AND CONCLUSIONS}
GRB 120326A is an unusual GRB, which shows a quite long plateau and a very late rebrightening in X-ray and optical bands at the same time. The similar shape in X-ray and optical bands of GRB 120326A suggests the same origin for them. The long plateau starts at several hundred seconds and ends at tens of thousands seconds. The peak time of the late rebrightening is about 30000 s. Because of such a curious phenomenon, a lot of telescopes did the follow-up observations of this GRB. We collect a lot of observational data, having early and late afterglow data to deepen our understanding of the energy injection mechanism.

Plateaus and rebrightenings/bumps in GRB afterglow light curves usually hint late reactivity of the GRB central engine (Dai \& Lu 1998; Zhang \& M\'{e}sz\'{a}ros 2001; Geng et al. 2013). However, it is rare that two phenomena take place in one GRB. In this study, we consider a new born millisecond pulsar with strong magnetic field as the central engine (Dai \& Liu 2012). Zhang and Yan (2011) also suggested that the central engine could produce a Poynting-flux dominated outflow. Due to the strong magnetic field and rapid rotation, the new born millisecond pulsar can release the energy comparable to that of the impulsive energy of the initial fireball. In this case, the very early afterglow may be directly powered by ultra-relativistic mildly magnetized outflows (Fan et al. 2004). It is also very possible that the continuous energy release by the pulsar will be injected into the external shock driven by the initial jet, which is currently the standard scenario for afterglow plateaus.

We analyze the energy injection model by means of numerical and analytical solutions, considering both the wind scenario and ISM scenario. The influence of the injection starting time, ending time, stellar wind density (or density of the circumburst environment), and injection luminosity on the shape of the afterglow light curves are explored. We find that the light curve of the afterglow is largely affected in the wind model by the parameters. We also find that although the light curve is slightly affected in the ISM case by these parameters, there is a significant ``bump'' at the late time only in the wind model.

For different GRBs, the circumburst environment (wind parameter $A_{*}$), the luminosity ($L_0$) and duration ($T_{\rm start}$, $T_{\rm end}$) of the energy injection may vary markedly. This may be one possible explanation to the observed diverse afterglow light curves. From Figure 3a and 4a, we show that the plateau and the late bump is an indicator of a wind type circumburst environment. GRB 120326A with such temporal behavior is an ideal template. Our results suggest that the circumburst environment of GRB 120326A is very likely a stellar wind. With reasonable parameter values, we give a good fitting to the X-ray and optical afterglow light curves of GRB 120326A (see Fig. 2). The model parameters from our fits are $A_{*}=0.45$, $T_{\rm start}=600$ s, $T_{\rm end}=3\times10^4$ s, $L_0=1.70\times10^{49}$ erg $\rm s^{-1}$.

\acknowledgments
We acknowledge the anonymous referee for useful comments and suggestions. We appreciate helpful discussion with Li-ping Xin, Da-bin Lin, and Ruo-Yu Liu. We acknowledge the use of the public data from the Swift, GCN Circulars archive and Fermi data archives. Our work also made use of data supplied by the UK Swift Science Data Centre at the University of Leicester. This work was supported by the National Basic Research Program of China (973 Program, Grant No. 2013CB834900 and No. 2014CB845800), the National Natural Science Foundation of China (Grants 11322328, 11233006, 11103015, 11033002, and U1331101). X. F. Wu acknowledges support by the One-Hundred-Talents Program, the Youth Innovation Promotion Association, and the Strategic Priority Research Program ``The Emergence of Cosmological Structures'' of the Chinese Academy of Sciences (Grant No. XDB09000000).

\begin{figure}\label{Fig_1}
\centering\includegraphics[angle=0,scale=1.0]{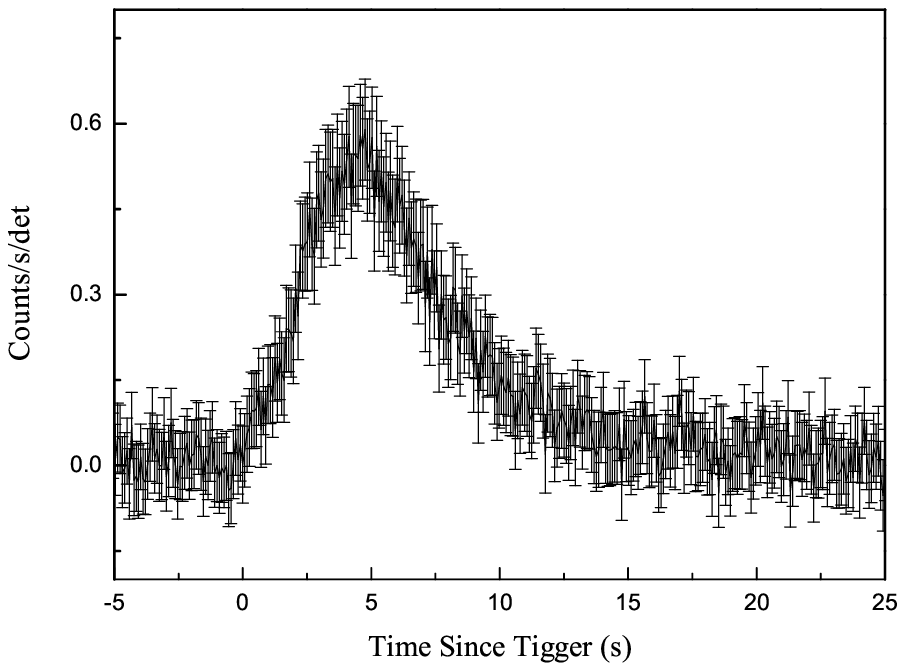}
\caption{{\em Left panel}: BAT count rate light curve of GRB 120326A. It shows the light curve is consisted with a single pulse. {\em Right panel}: Time-integrated spectrum of GRB 120326A, which is derived from the Fermi NaI 1 and NaI 2 data. The line is our best fit by using the Band function.}
\end{figure}

\begin{figure}\label{Fig_2}
\centering\includegraphics[angle=0,scale=1.5]{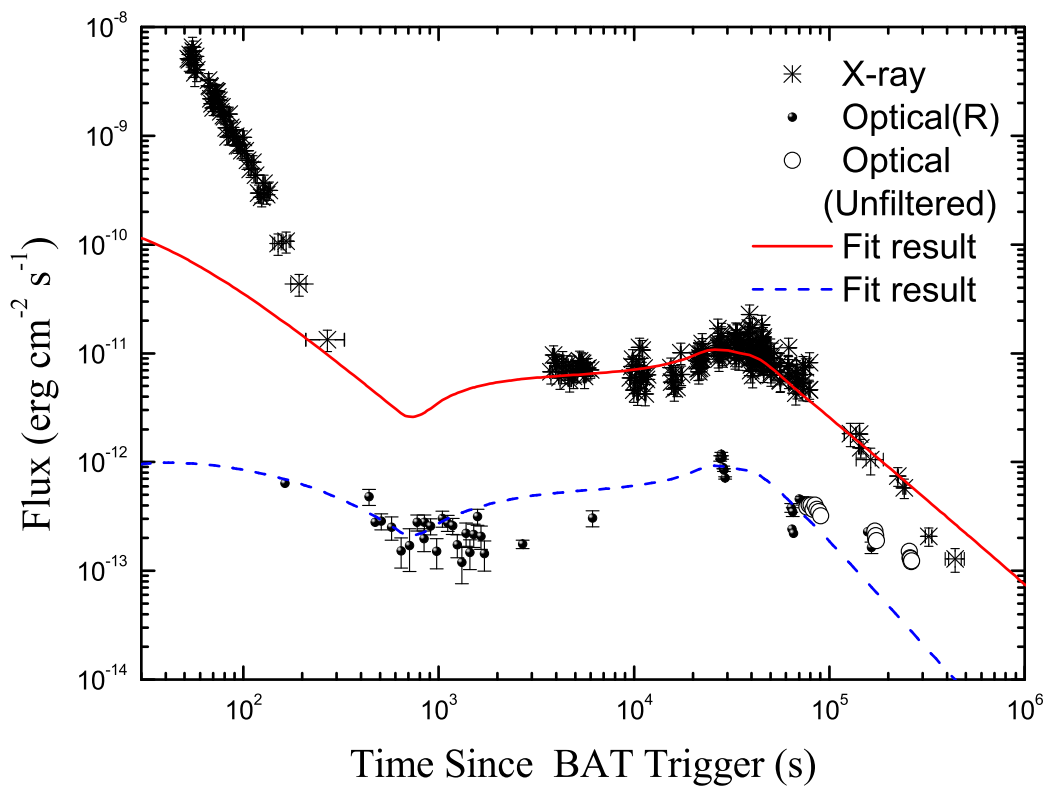}
\caption{The afterglow light curves of GRB 120326A. The black asterisks correspond to the $\it Swift/XRT$ data. The black dots represent the $R$ and $r$ bands optical data. The black open circles show unfiltered optical. All are collected from GCN. The red and blue lines are the best fitting lines with the energy injection model. The $R$ and $r$ bands data are used to limit the model and the unfiltered data are only used for reference.}
\end{figure}

\begin{figure}\label{Fig_3}
\centering\includegraphics[angle=0,scale=0.8]{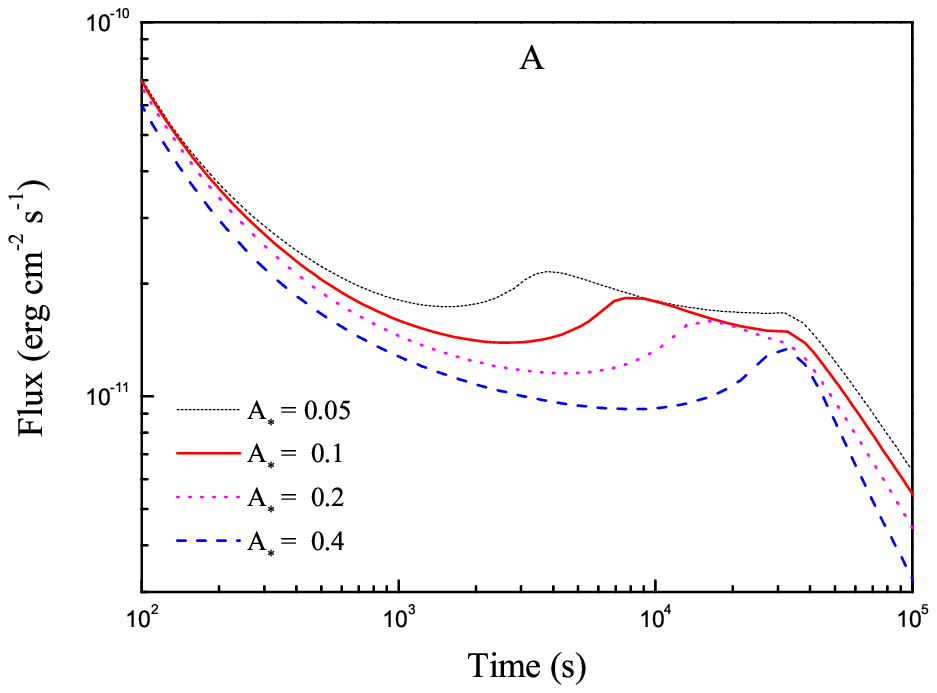}
\centering\includegraphics[angle=0,scale=0.8]{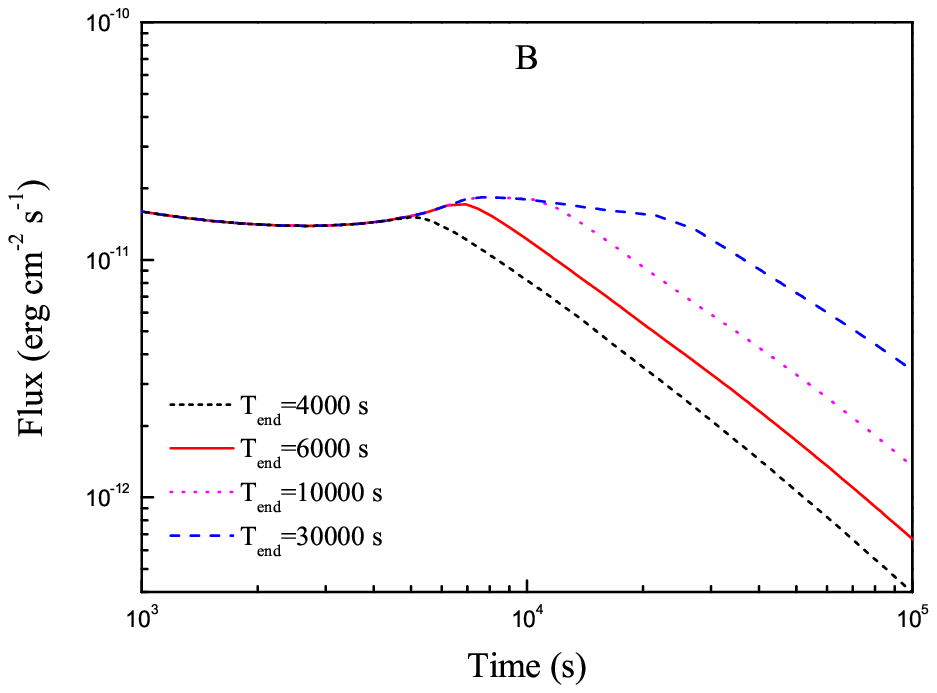}
\centering\includegraphics[angle=0,scale=0.8]{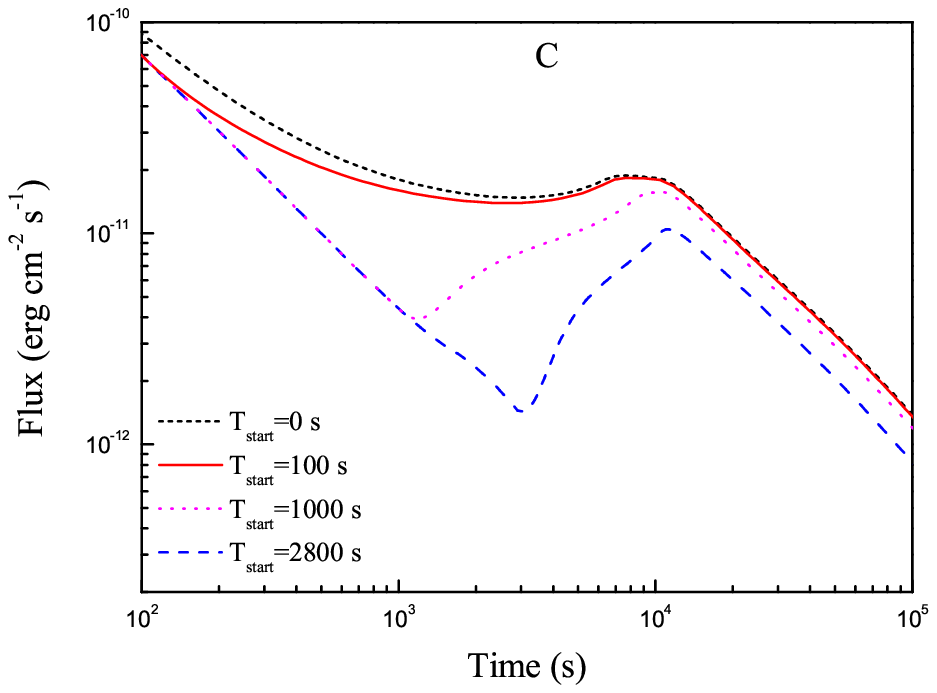}
\centering\includegraphics[angle=0,scale=0.8]{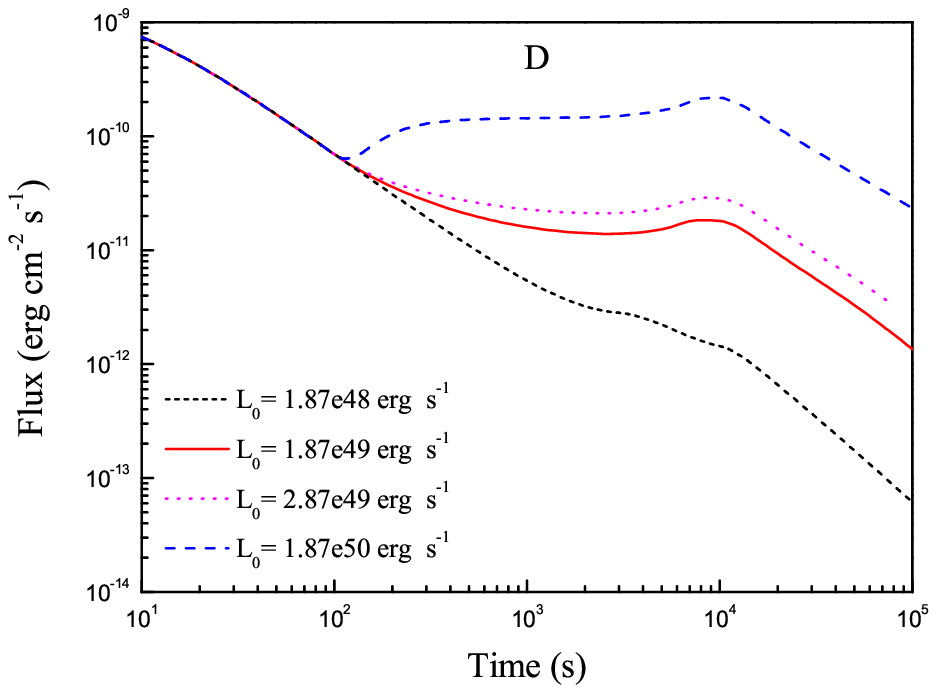}
\caption{Effects of various parameters on the X-ray afterglow light curve in the wind model. Panels a, b, c and d show the effects of the wind parameter $A_{*}$, the starting time $T_{\rm start}$ and ending time $T_{\rm end}$ of energy injection, and the injection luminosity $L_{0}$, respectively. In our calculations, the standard choice of the values of these parameters are $A_{*}=0.1$, $T_{\rm start}=100$ s,  $L_0=1.87\times10^{49}$ erg $\rm s^{-1}$, and $T_{\rm end}=10$ ks (the only exception is when investigating the effect of $A_{*}$, $T_{\rm end}=30$ ks is adopted).}
\end{figure}
\begin{figure}\label{Fig_4}
\centering\includegraphics[angle=0,scale=0.8]{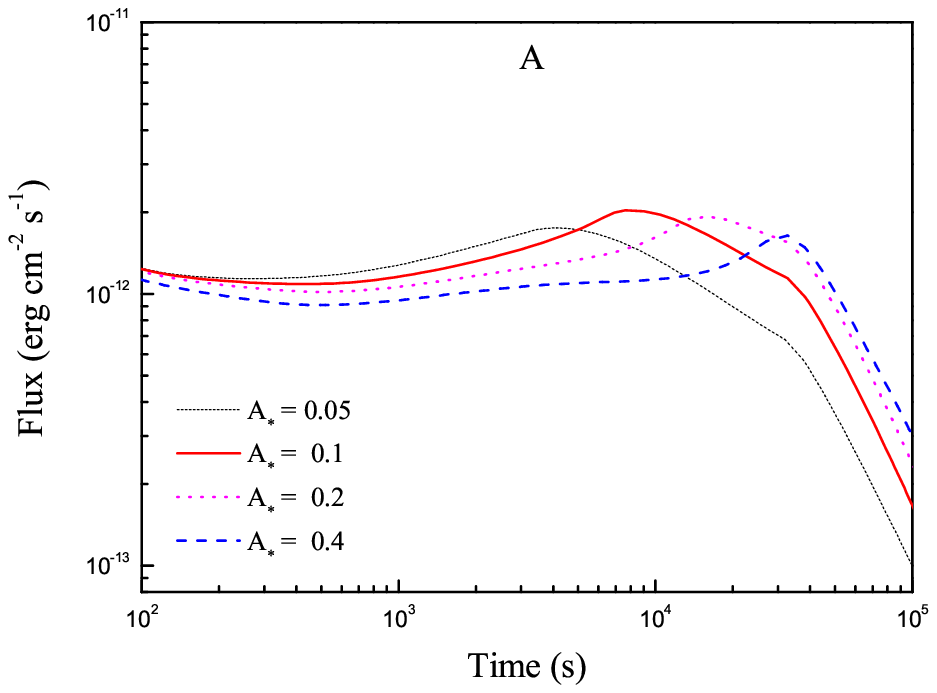}
\centering\includegraphics[angle=0,scale=0.8]{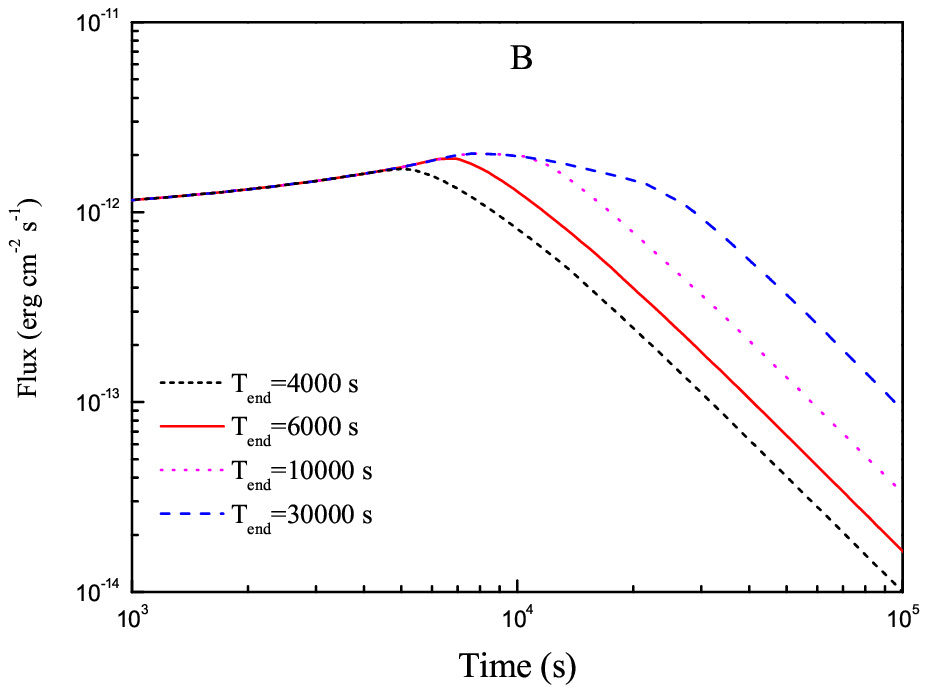}
\centering\includegraphics[angle=0,scale=0.8]{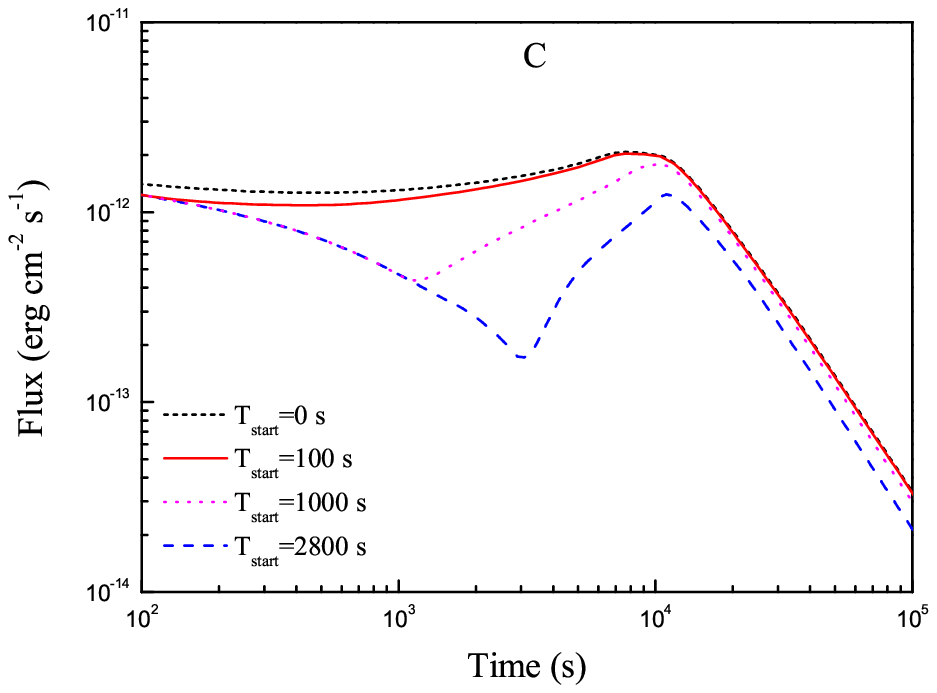}
\centering\includegraphics[angle=0,scale=0.8]{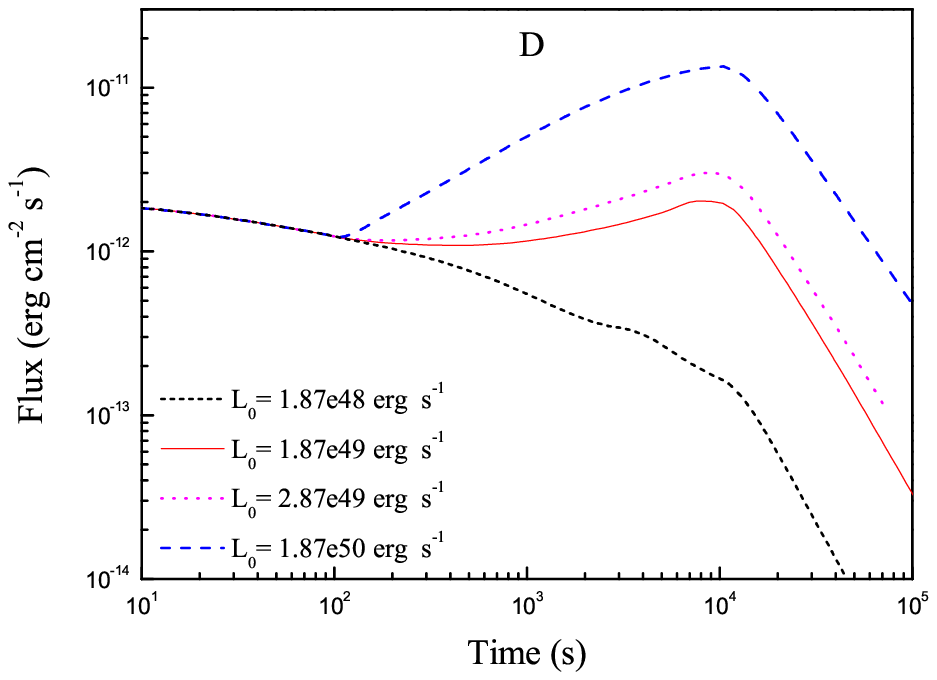}
\caption{ Effects of various parameters on the optical afterglow light curve in the wind model. Panels a, b, c and d show the effects of the wind parameter $A_{*}$, the starting time $T_{\rm start}$ and ending time $T_{\rm end}$ of energy injection, and the injection luminosity $L_{0}$, respectively. In our calculations, the standard choice of the values of these parameters are $A_{*}=0.1$, $T_{\rm start}=100$ s, $L_0=1.87\times10^{49}$ erg $\rm s^{-1}$, and $T_{\rm end}=10$ ks (the only exception is when investigating the effect of $A_{*}$, $T_{\rm end}=30$ ks is adopted).}
\end{figure}

\begin{figure}\label{Fig_5}
\centering\includegraphics[angle=0,scale=0.8]{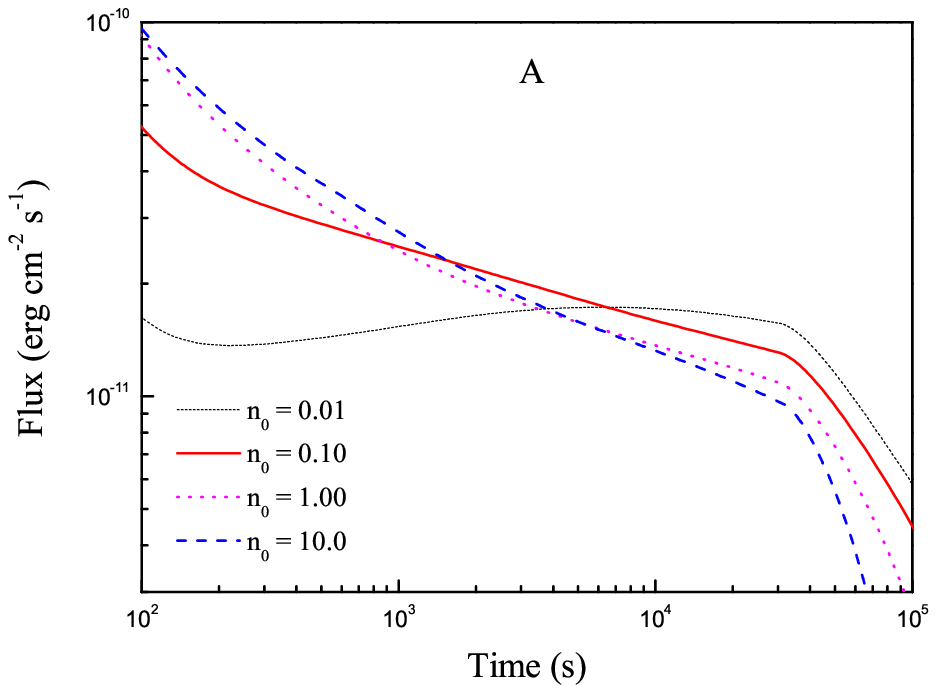}
\centering\includegraphics[angle=0,scale=0.8]{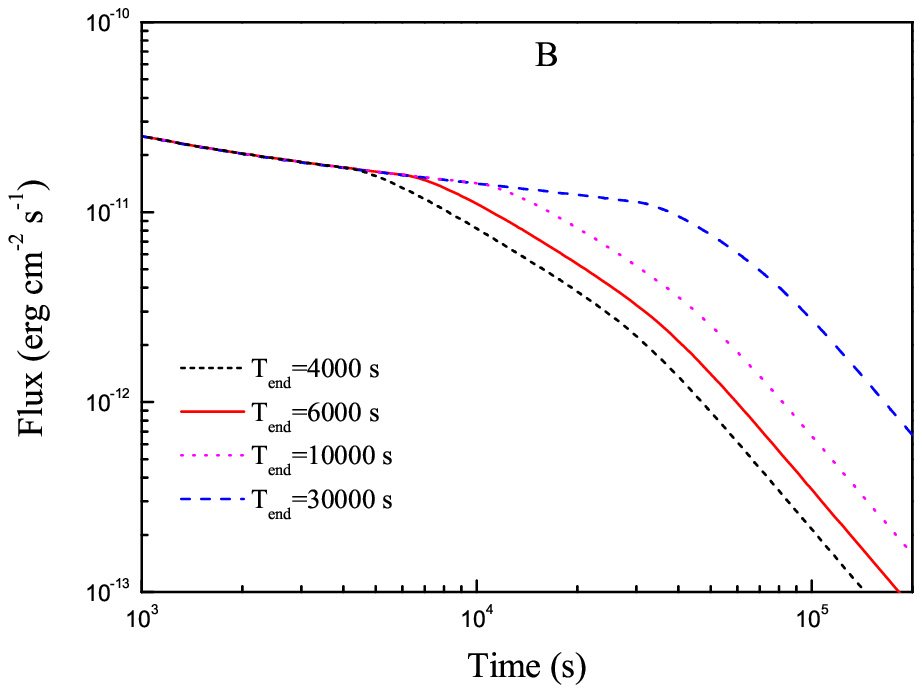}
\centering\includegraphics[angle=0,scale=0.8]{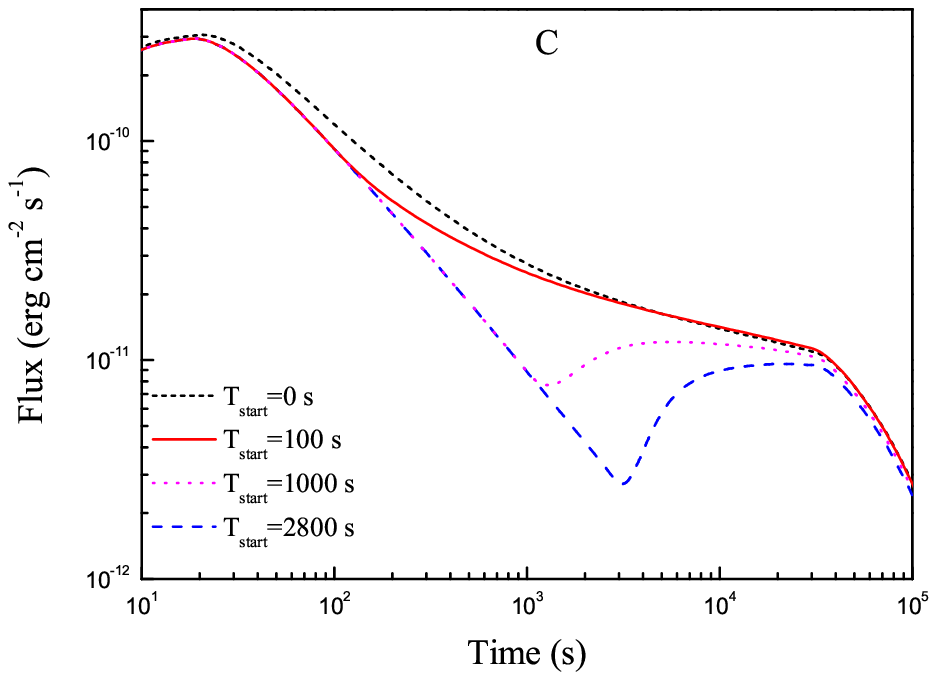}
\centering\includegraphics[angle=0,scale=0.8]{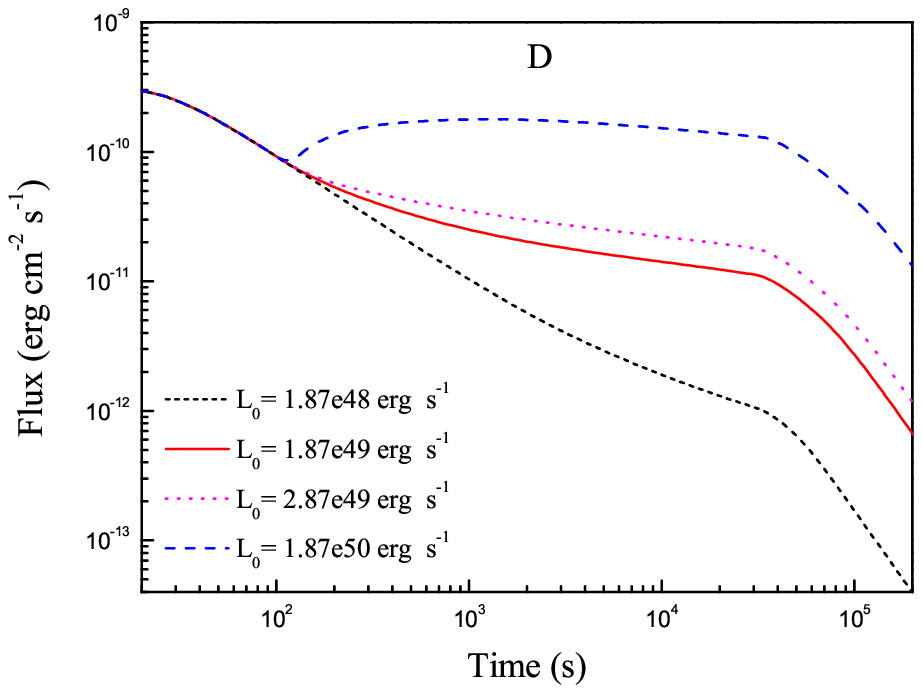}
\caption{Effects of various parameters on the X-ray afterglow light curve in the ISM model. Panels a, b, c and d show the effects of the density of the circumburst environment $n_{0}$, the starting time $T_{\rm start}$ and ending time $T_{\rm end}$ of energy injection, and the injection luminosity $L_{0}$, respectively. In our calculations, the standard choice of the parameters are $n_{0}=1.0$ $\rm cm^{-3}$, $T_{\rm start}=100$ s, $T_{\rm end}= 30$ ks, and $L_0=1.87\times10^{49}$ erg $\rm s^{-1}$.}
\end{figure}
\begin{figure}\label{Fig_6}
\centering\includegraphics[angle=0,scale=0.8]{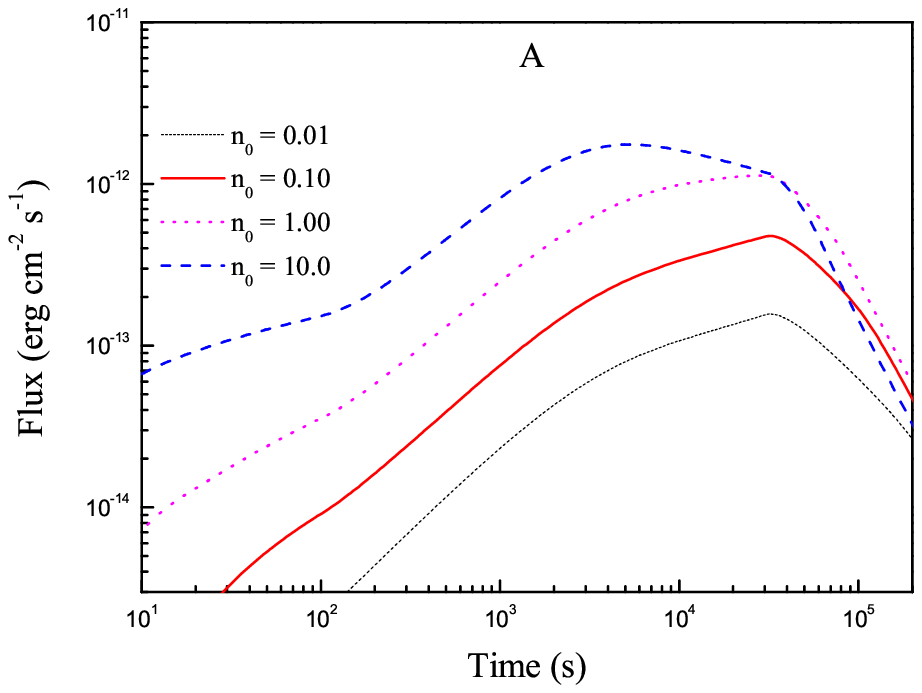}
\centering\includegraphics[angle=0,scale=0.8]{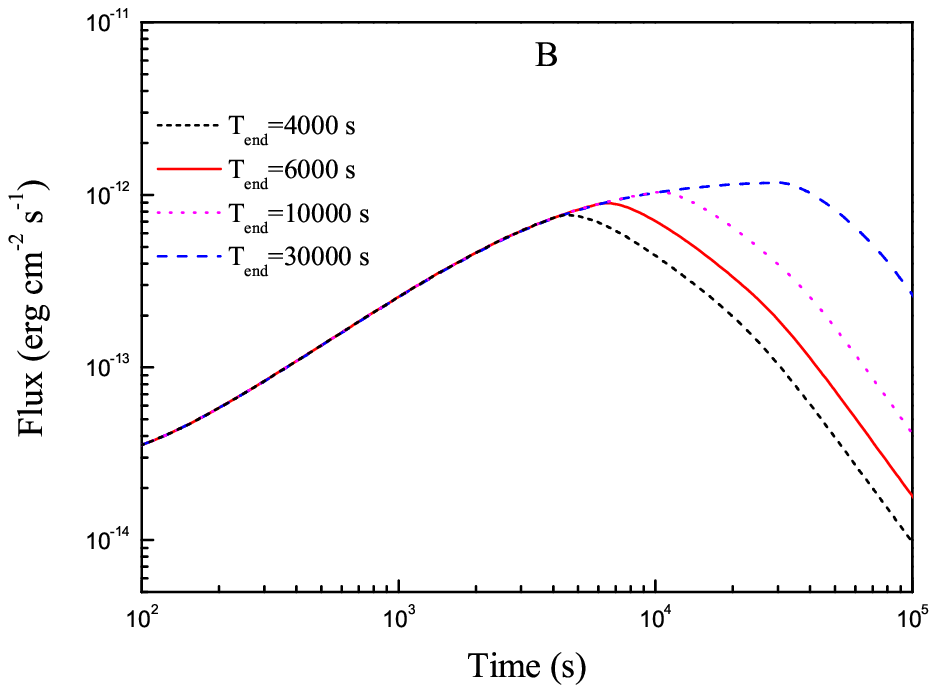}
\centering\includegraphics[angle=0,scale=0.8]{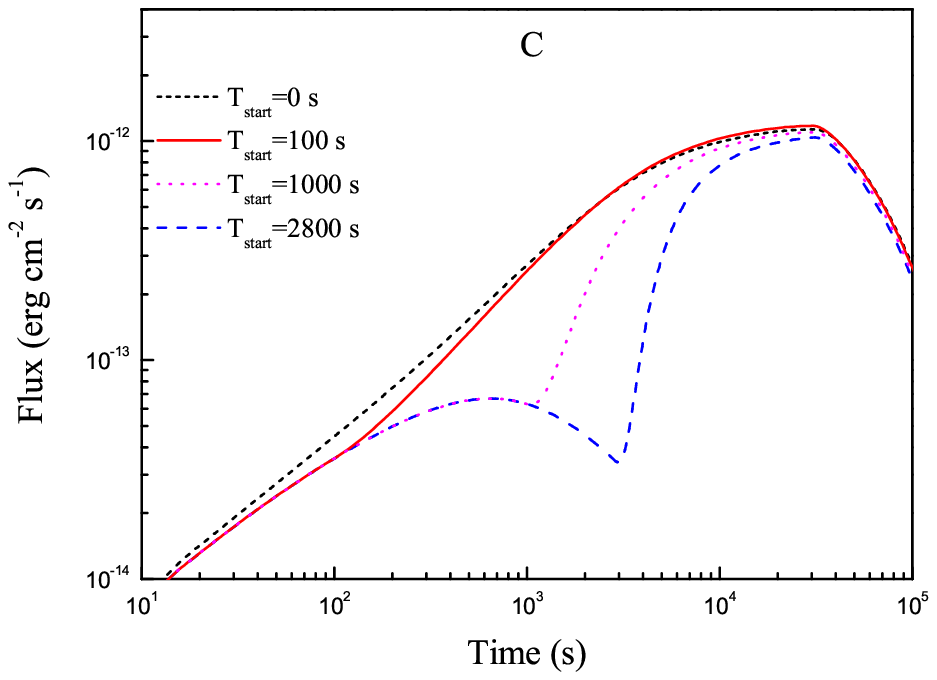}
\centering\includegraphics[angle=0,scale=0.8]{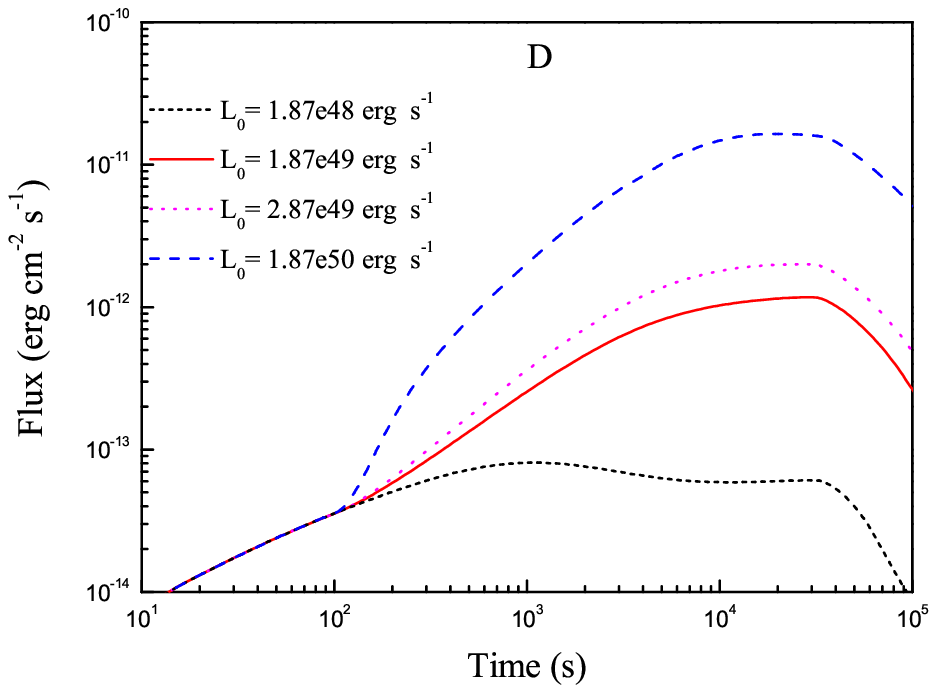}
\caption{Effects of various parameters on the optical afterglow light curve in the ISM model. Panels a, b, c and d show the effects of the density of the circumburst environment $n_{0}$, the starting time $T_{\rm start}$ and ending time $T_{\rm end}$ of energy injection, and the injection luminosity $L_{0}$, respectively. In our calculations, the standard choice of the parameters are $n_{0}=1.0$ $\rm cm^{-3}$, $T_{\rm start}=100$ s, $T_{\rm end}= 30$ ks, and $L_0=1.87\times10^{49}$ erg $\rm s^{-1}$.}
\end{figure}

\begin{figure}\label{Fig 7}
\centering\includegraphics[angle=0,scale=0.8]{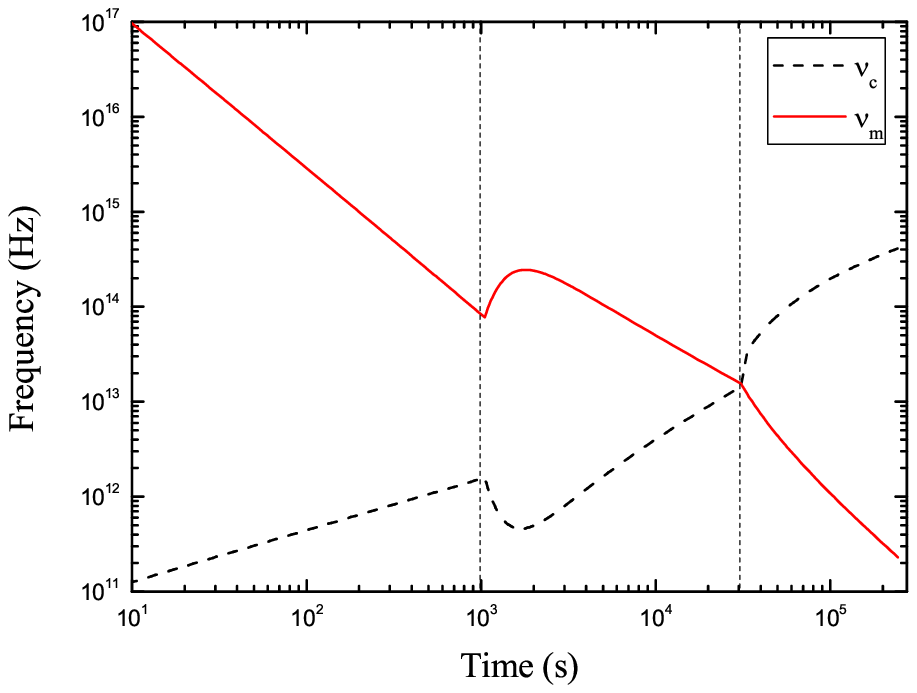}
\centering\includegraphics[angle=0,scale=0.8]{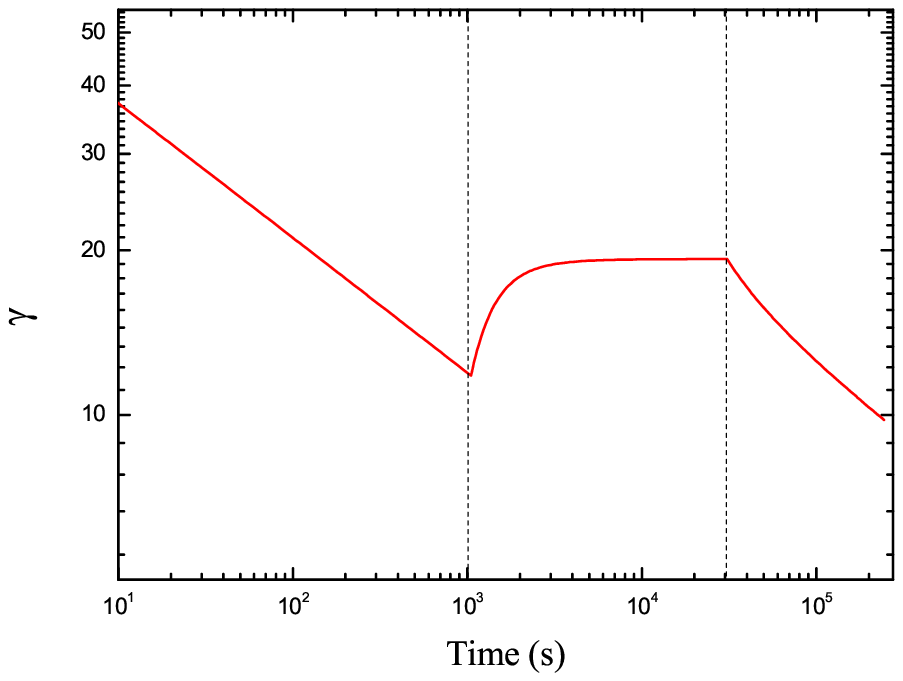}
\caption{Evolution of $\nu_{c}$ (dashed line), $\nu_{m}$ (solid line) and $\gamma$ with time for the case of GRB 120326A afterglow. Two vertical dashed lines indicate the starting time and ending time of energy injection, respectively.}
\end{figure}

\begin{deluxetable}{cccccc}\label{table_1}
\tabletypesize{\scriptsize}
\tablewidth{350pt}
\tablecaption{GRB 120326A optical observations collected from GCN$^{1}$. }
\tablehead{\colhead{Time} &\colhead{Exp.Time} &\colhead{Magnitude} &\colhead{Mag Error} &\colhead{$\rm Filters$} &\colhead{$\rm Ref^{2}$}\\
\colhead{(s)} &\colhead{(s)} &\colhead{(mag)} &\colhead{(mag)} &\colhead{} &\colhead{}}
\startdata
163	&	59.4	&	18.2	&		&	R	&	1	\\
473	&	945	&	19.1	&		&	R	&	1	\\
846	&	30	&	19.1	&	0.2	&	r	&	2	\\
2688	&	120	&	19.6	&	0.1	&	r	&	2	\\
440	&	60	&	18.51	&	0.2	&	R	&	3	\\
507	&	60	&	19.07	&	0.2	&	R	&	3	\\
574	&	60	&	19.21	&	0.3	&	R	&	3	\\
641	&	60	&	19.75	&	0.4	&	R	&	3	\\
708	&	60	&	19.63	&	0.6	&	R	&	3	\\
775	&	60	&	19.1	&	0.2	&	R	&	3	\\
842	&	60	&	19.48	&	0.3	&	R	&	3	\\
909	&	60	&	19.19	&	0.2	&	R	&	3	\\
976	&	60	&	19.77	&	0.4	&	R	&	3	\\
1043	&	60	&	19.01	&	0.2	&	R	&	3	\\
1110	&	60	&	19.11	&	0.2	&	R	&	3	\\
1177	&	60	&	19.18	&	0.2	&	R	&	3	\\
1244	&	60	&	19.62	&	0.3	&	R	&	3	\\
1311	&	60	&	20.02	&	0.5	&	R	&	3	\\
1379	&	60	&	19.35	&	0.3	&	R	&	3	\\
1446	&	60	&	19.79	&	0.4	&	R	&	3	\\
1513	&	60	&	19.38	&	0.3	&	R	&	3	\\
1580	&	60	&	18.97	&	0.2	&	R	&	3	\\
1647	&	60	&	19.42	&	0.3	&	R	&	3	\\
1714	&	60	&	19.82	&	0.4	&	R	&	3	\\
6125	&	720	&	19	&	0.2	&	unfiltered 	&	4	\\
63720	&	300	&	18.7	&	0.1	&	R	&	5	\\
64800	&	300	&	18.8	&	0.1	&	R	&	5	\\
72720	&	3600	&	18.6	&	0.1	&	R	&	6	\\
70020	&	300	&	18.56	&		&	R	&	7	\\
27639	&	300	&	17.63	&	0.06	&	R	&	8	\\
27964	&	300	&	17.53	&	0.04	&	R	&	8	\\
28282	&	300	&	17.61	&	0.04	&	R	&	8	\\
28601	&	300	&	17.84	&	0.03	&	R	&	8	\\
28915	&	300	&	17.9	&	0.03	&	R	&	8	\\
29239	&	300	&	18.08	&	0.03	&	R	&	8	\\
157320	&	300	&	19.32	&		&	R	&	9	\\
164070	&	1500	&	19.68	&	0.07	&	R	&	10	\\
76781	&	120	&	18.75	&		&	unfiltered 	&	11	\\
78675	&	120	&	18.7	&		&	unfiltered 	&	11	\\
80570	&	120	&	18.7	&		&	unfiltered 	&	11	\\
82468	&	120	&	18.8	&		&	unfiltered 	&	11	\\
84363	&	120	&	18.7	&		&	unfiltered 	&	11	\\
86260	&	120	&	18.8	&		&	unfiltered 	&	11	\\
88158	&	120	&	18.85	&		&	unfiltered 	&	11	\\
90244	&	120	&	18.95	&		&	unfiltered 	&	11	\\
170707	&	120	&	19.3	&		&	unfiltered 	&	11	\\
172603	&	120	&	19.48	&		&	unfiltered 	&	11	\\
172604	&	120	&	19.41	&		&	unfiltered 	&	11	\\
174501	&	120	&	19.52	&		&	unfiltered 	&	11	\\
256864	&	120	&	19.77	&		&	unfiltered 	&	11	\\
258667	&	120	&	19.91	&		&	unfiltered 	&	11	\\
260619	&	120	&	19.93	&		&	unfiltered 	&	11	\\
262582	&	120	&	20	&		&	unfiltered 	&	11	\\
263298	&	120	&	19.98	&		&	unfiltered 	&	11	\\
\enddata
\tablenotetext{1}{From left to right: Time since burst, Exposure Time, Magnitude, Magnitude Error, Filters, and References.}
\tablenotetext{2}{References: (1) Klotz et al. 2012b; (2) Guidorzi 2012; (3) Dintinjana \& Mikuz 2012; (4) Hentunen et al. 2012; (5) Zhao et al. 2012; (6) Soulier 2012; (7) Xin et al. 2012a; (8) Jang et al. 2012; (9) Xin et al. 2012b; (10) Sahu et al. 2012; (11)Quadri et al. 2012.}
\end{deluxetable}

\end{document}